\documentstyle[12pt]{article}
\def\be {\begin{equation}}
\def\ee {\end{equation}}
\def\ba {\begin{eqnarray}}
\def\ea {\end{eqnarray}}

\def\bi {\begin{itemize}}
\def\ei {\end{itemize}}
\begin{document}
\def\bea{\begin{eqnarray}}
\def\eea{\end{eqnarray}}
\title{\bf {The Self-Gravitational Corrections as the Source for
Stiff Matter on the Brane in $SAdS_5$ Bulk }}
 \author{M.R. Setare  \footnote{E-mail: rezakord@ipm.ir}
  \\{Physics Dept. Inst. for Studies in Theo. Physics and
Mathematics(IPM)}\\
{P. O. Box 19395-5531, Tehran, IRAN }}
\date{\small{}}

\maketitle
\begin{abstract}
A $D3-$brane with non-zero energy density is considered as the
boundary of a five dimensional Schwarzschild anti de Sitter bulk
background. Taking into account the semi-classical corrections to
the black hole entropy that arise as a result of the
self-gravitational effect, and employing the AdS/CFT
correspondence, we obtain the self-gravitational correction to the
first Friedmann-like equation. The additional term in the Hubble
equation due to the self-gravitational effect goes as $a^{-6}$.
Thus, the self-gravitational corrections act as a source of stiff
matter contrary to standard FRW cosmology where the charge of the
black hole plays this role.
 \end{abstract}

\newpage

\section{Introduction}
In recent years, a lot of interest has been raised in studying the
cosmology of universes with extra dimensions
\cite{{rs1},{rs2},{add}}. The common feature of all these models
is the distinction of observable universe (the three-brane )from
the rest of the universe (the bulk). Ordinary matter field are
assumed to live on the brane while gravity propagates in the whole
spacetime.  The braneworld scenario gained momentum as a solution
to the hierarchy problem \cite{{rs1},{rs2},{add}}. The holographic
principle, meanwhile, was first realized in string theory via the
AdS/CFT correspondence\cite{{mal},{wit1},{gab}}.  The essence of
braneworld holography {\cite{heb}-\cite{set}} can be captured in
the following claim : Randall-Sundrum branworld gravity is dual to
a CFT with a UV cutoff, coupled to gravity on the brane. Formal
evidence for this claim was provided by studying a brane universe
in the background of the Schwarzschild-AdS Bblack hole. The
introduction of the black hole on the gravity side of the AdS/CFT
correspondence corresponds to considering finite temperature
states in the dual CFT \cite{wit2}. In the context of braneworld
holography, Savonije and Verlinde demonstrated that their induced
braneworld cosmology could alternatively be described as the
standard FRW cosmology driven by the energy density of this dual
CFT \cite{{verlinde},{savonije}}.\\
Black hole thermodynamic quantities depend on the Hawking
temperature via the usual thermodynamic relations. The Hawking
temperature undergoes corrections from many sources: the quantum
corrections \cite{kaul}- \cite{set6}, the self-gravitational
corrections \cite{{KKW},{corrections}}, and the corrections due to
the generalized uncertainty principle \cite{gupdas, set9}.
Concerning the quantum process called Hawking effect
\cite{hawking1} much work has been done using a fixed background
during the emission process. The idea of Keski-Vakkuri, Kraus and
Wilczek (KKW) \cite{KKW} is to view the black hole background as
dynamical by treating the Hawking radiation as a tunnelling
process. The energy conservation is the key to this description.
The total (ADM) mass is kept fixed while the mass of the black
hole under consideration decreases due to the emitted radiation.
The effect of this modification gives rise to additional terms in
the formulae concerning the known results for black holes
\cite{corrections}; a nonthermal partner to the thermal spectrum
of the Hawking radiation shows up.\\
In the present paper, we take into account corrections to the
entropy of the  five-dimensional Schwarzschild- anti de Sitter
black hole (abbreviated to $SAdS_5$ in the sequel) that arise due
to the self-gravitational effect. Previous studies of the
Cardy-Verlinde formula (or the corresponding Friedmann equation)
in an AdS/CFT context have attracted a lot of attention
\cite{{setelias1},{setelias2}}. In the previous paper
\cite{setare}, by studying the case of an empty brane with zero
cosmological constant, embedded in an $SAdS_5$ black hole bulk
spacetime, and introducing the self-gravitational correction to
the Cardy-Verlinde formula, we were able to find a host of
interesting cosmological solutions for the brane universe. The
self-gravitational correction, acts as a source for stiff matter
on the brane, whose equation of state is simply given by the
pressure being equal to the energy density. Due to the
self-gravitational corrections, a bouncing universe could, arise,
i.e. a universe that bounce from a contracting phase to an
expanding one without encountering a singularity.\\
Previously have been shown that the charged Ads/CFT black hole
background provides support for a singularity-free cosmology
\cite{kan}-\cite{mu} in which the big Bang singularity is not
present. In the present paper we consider the more physically
relevant case in which a perfect fluid with equation of state of
radiation is present on the brane.

\section{Self-Gravitational Corrections to FRW Brane Cosmology}

In the asymptotic coordinates, the $SAdS_5$ black hole metric is
 \bea
ds^{2}=-F(r)dt^{2}+\frac{1}{F(r)}dr^{2}+r^{2} d\Omega_{(3)}^{2},
\label{metric1} \eea where \be
F(r)=1-\frac{\mu}{r^{2}}+r^{2},\label{fterm} \ee and $l=1$ is the
$AdS$ radius . The parameter $\mu$ is
proportional to the ADM mass $M$ of the black hole.\\
Due to the self-gravitational corrections, the modified
Cardy-Verlinde formula for the entropy of the $SAdS_{5}$ black
hole is given as \cite{setare} \be S_{CFT}=\frac{2\pi
r}{3}\sqrt{\Bigg| \left[E_{C}-\frac{1}{r}\omega\right]
\left[\left(2E_{4}-E_{C}\right)-\frac{1}{r}\omega\right]\Bigg|}
\hspace{1ex} \ee and keeping terms up to first order in the
emitted energy $\omega$, it takes the form

\be S_{CFT}=s_{CFT}\left(1-\varepsilon \omega\right) \hspace{1ex}
\label{modS} \ee where the small parameter $\varepsilon$ is given
by \be
\varepsilon=\frac{1}{r}\frac{E_{4}}{E_{C}\left(2E_{4}-E_{C}\right)}
\hspace{1ex}. \label{epsilon} \ee where $E_C$ is the Casimir
energy, the four-dimensional energy $E_4$, is given by
 \bea E_4=\frac{l}{r}E \label{ener2} \eea where $E$ is the
 thermodynamical energy of the black hole.\\
 We now consider a $4$-dimensional brane in the
$SAdS_5$ black hole background. This $4$-dimensional brane can be
regarded as the boundary of the $5$-dimensional $SAdS_{5}$ bulk
background. Let us first replace the radial coordinate $r$ with
$a$ and so the line element (\ref{metric1}) \bea
ds^{2}=-F(a)dt^{2}+\frac{1}{F(a)}da^{2}+a^{2}d\Omega_{(3)}^{2},
\label{met} \eea Within the context of the AdS/CFT correspondence,
Savonije and Verlinde studied the CFT/FRW-cosmology relation from
the Randall-Sundrum type braneworld perspective \cite{savonije}.
They showed that the entropy formulas of the CFT coincides with
the Friedmann equations when the brane crosses the black hole
horizon.

In the case of a $4$-dimensional timelike
 \be
ds_{(4)}^{2}=-d\tau^{2}+a^{2}(\tau)d\Omega_{(3)}^{2} \hspace{1ex},
\ee One of the identifications that supports the CFT/FRW-cosmology
relation  is \be H^{2}=\left(\frac{2G_4}{V}\right)^{2}{S}^{2}
\label{huble} \ee where $H$ is the Hubble parameter defined by
$H=\frac{1}{a}\frac{da}{d\tau}$ and V is the volume of the
$3$-sphere ($V=a^{3}V_{3}$). The $4$-dimensional Newton constant
$G_4$ is related to the $5-$dimensional one $G_5$ by \be
 G_{4}=\frac{2}{l}G_{5}
 \hspace{1ex}.
\ee It was shown that at the moment that the $4$-dimensional
timelike brane crosses the cosmological horizon, i.e. when
$a=a_{b}$, the CFT entropy and the entropy of the $SAdS_5$ black
hole are identical. By substituting (\ref{modS}) into
(\ref{huble}), we obtain the self-gravitational corrections to
the motion of the CFT-dominated brane \be
H^{2}=\left(\frac{2G_4}{V}\right)^{2}S_{CFT}^{2}\left(1-\varepsilon
\omega\right)^{2} \label{modH} \hspace{1ex}. \ee It is obvious
that from the first term on the right-hand side of (\ref{modH})
we get the standard Friedmann equation with the appropriate
normalization \be H^{2}=\frac{-k}{a^{2}}+\frac{8\pi G_{4}}{3}\rho
\label{shub} \ee where $\rho$ is the energy density defined by
$\rho=E_{4}/V$, and $k$ taking values ${+1,0,-1}$ in order to
describe, respectively, the elliptic, flat, and hyperbolic
horizon geometry of the $SAdS_5$ bulk black hole. If we consider
the more physically relevant case in which a perfect fluid with
equation of state of radiation is present on the brane, then the
first Friedmann equation takes the following form \be
H^{2}=\frac{-k}{a^{2}}+\frac{8\pi
G_{4}}{3}\rho-\frac{1}{l^{2}}+\frac{4\pi}{3M_{p}^{2}\rho_{0}}(\rho_{0}+\rho_{br})^{2}
\label{shab1} \ee where $\rho_{0}$ is the tension of the brane,
while $\rho_{br}$ is the energy density of radiation. The Hubble
equation can be rewritten as \be H^{2}=\frac{-k}{a^{2}}+\frac{8\pi
G_{4}}{3}\rho+\frac{\Lambda_{4}}{3}+\frac{8\pi}{3M_{p}^{2}}(\frac{\rho_{br}^{2}}{2\rho_{0}}+\rho_{br})
\label{shab2} \ee where \be \Lambda_{4}=\frac{4\pi
\rho_{0}}{M_{p}^{2}}-\frac{3}{l^{2}}=\Lambda_{br}-\frac{3}{l^{2}}
\label{cosmo} \ee is the effective cosmological constant of the
brane. The correction to the FRW equation due to the
self-gravitation effect is expressed by the second term in the
right-hand side of Eq.(\ref{modH}). Keeping terms up to first
order in the emitted energy $\omega$, the modified Hubble
equation due to the self-gravitation correction is  \be
H^{2}=\frac{-k}{a^{2}}+\frac{8\pi
G_{4}}{3}\rho+\frac{\Lambda_{4}}{3}+\frac{8\pi}{3M_{p}^{2}}(\frac{\rho_{br}^{2}}{2\rho_{0}}+\rho_{br})
 - \frac{8\pi
G_{4}}{3}\left[\frac{4\pi
G_{4}}{3}\frac{1}{a^{2}V_{3}}\rho\right]\omega \label{gmodHubble}
\ee  By tuning the bulk cosmological constant and the brane
tension $\Lambda_{br}$, the effective four dimensional
cosmological constant $\Lambda_{4}$ can be set to zero, here we
would like consider this critical brane. Also for the very small
$a$ the curvature term $\frac{-k}{a^{2}}$ can be neglected in the
above equation relative to the other contributors. The radiation
energy density $\rho_{br}$ on the brane is as \be
\rho_{br}=\frac{\rho_{r}}{a^{4}} \label{tens} \ee where
$\rho_{r}$ is a constant, then we can rewrite the cosmological
equation (\ref{gmodHubble}) as \be H^{2}=(\epsilon_{3}M+\frac{8\pi
\rho_{r}}{3M_{p}^{2}})\frac{1}{a^4}-\frac{\epsilon_{3}^{2}M\omega}{2a^6}+\frac{4\pi
\rho_{r}^{2} }{3M_{p}^{2}\rho_{0}a^{8}} \label{modhub2}\ee where
\be \epsilon_{3}=\frac{16\pi G_5}{3V_3} \ee The evolution of the
system can be solved exactly, as one can most simply realize by
using conformal time $\eta$, defined as $dt=a(\eta)^{2}d\eta$ and
new variable $x=a^2$: \be \frac{1}{4}x'^{2}=c_1x^{2}+c_2x+c_3
\label{deeq} \ee where \be c_1=\epsilon_{3}M+\frac{8\pi
\rho_{r}}{3M_{p}^{2}} \label{c1}\ee \be
c_2=\frac{-\epsilon_{3}^{2}M\omega}{2} \label{c2} \ee \be
c_3=\frac{4\pi\rho_{r}^{2} }{3M_{p}^{2}\rho_{0}} \label{c3} \ee
With the assumption that a bounce does indeed take place, and
setting $x'=0$, the following condition emerges \be
\frac{\epsilon_{3}M^{2}\omega^{2}}{4}\geq \frac{16
\pi\rho_{r}^{2} }{3M_{p}^{2}\rho_{0}}(\epsilon_{3}M+\frac{8\pi
\rho_{r}}{3M_{p}^{2}}) \label{condi} \ee The solution of
Eq.(\ref{deeq}) is as \be
a^{2}=\frac{-c_2}{2c_1}+\sqrt{\Delta}\cosh(2\sqrt{c_1}\eta)
\label{scal} \ee where \be
\Delta=\frac{-c_3}{c_1}+(\frac{c_2}{c_1})^{2}>0. \label{delta}\ee
The time variables are related as \be t=
\frac{-c_2}{2c_1}\eta+\sqrt{\frac{\Delta}{4c_1}}\sinh(2\sqrt{c_1}\eta)
\label{time} \ee We set $\eta=0$ at the bounce, then $t$ is also
vanish at the bounce. Therefore the minimal value of the scale
factor is given by \be
a_{min}^{2}=\frac{-c_2}{2c_1}+\sqrt{\Delta}.\label{mini} \ee An
immediate consequence of last term in Eq.(\ref{modhub2}) is that
the bounce can only occur if self-gravitational corrections
satisfy condition (\ref{condi}). In the other hand, for an empty
brane, a bounce obtain for any value of the self-gravitational
corrections.
\section{Conclusions}
In this paper we have considered  a four-dimensional timelike
brane with non-zero energy density as the boundary of the
$SAdS_{5}$ bulk background. Exploiting the CFT/FRW-cosmology
relation, we have derived the self-gravitational corrections to
the first Friedmann-like equation which is the equation of the
brane motion. The additional term that arises due to the
semiclassical analysis, can be viewed as stiff matter where the
self-gravitational corrections act as the source for it. This
result is contrary to standard analysis that regards the charge
of $SAdS_{5}$ bulk black hole as the source for stiff matter. In
previous paper \cite{setare} we have considered an empty critical
brane. The effective cosmological constant $\Lambda_{4}$ on the
brane gets a contribution from the bulk cosmological constant and
the brane tension. For the critical brane we finely tune
$\Lambda_{4}$ to zero. In the non-critical empty brane case, the
solution of FRW equation at very small value of the scale factor,
are close to the behaviour for critical brane case. The
cosmological constant term dominant at large value of scale
factor, where the stiff matter term becomes irrelevant. In this
paper we have focused on the critical non-empty brane case. This
is more physically relevant case in which a perfect fluid with
equation of state of radiation is present on the brane. For small
scale factor $a$ we have neglected the curvature term
$\frac{-k}{a^{2}}$. Similar to the empty brane case even if we
consider the non-zero cosmological constant term (non-critical
brane case) for the small scale factor, this is irrelevant. The
first Friedmann equation gets a contribution proportional to
$a^{-6}$ due to the self-gravitational corrections with an
unconventional negative sign. This term behave like the stiff
matter on the brane. It is obvious that it is dominant at early
time of the brane evolution, also have interesting cosmological
consequences. The bounce can be attributed to the negative-energy
matter, which dominates at small values of $a$ and create a
significant enough repulsive force so that a big crunch is
avoided.\\
In the empty brane case, a bounce will be obtained due to the
self-gravitational corrections, regardless of how small, but in
the non-empty brane case the bounce can only occur if the
self-gravitational correction satisfies the condition
(\ref{condi}).

\end{document}